\documentclass[amsmath,amssymb,nofootinbib,showpacs,twocolumn,nofootinbib]{revtex4-1}
\usepackage{dcolumn}
\usepackage{bm}
\usepackage{graphicx}
\usepackage{epstopdf}
\usepackage{subfig}
\usepackage{epsfig}
\usepackage{color}
\usepackage{wasysym}
\usepackage{natbib}
\usepackage{twoopt}
\usepackage{dashrule}
\usepackage{enumerate}
\usepackage[breaklinks=true,colorlinks=true,linkcolor=blue,citecolor=blue,urlcolor=blue,pdfauthor={JieFeng},pdftitle={combinedFit}]{hyperref}
\usepackage{booktabs}
\usepackage{ulem}

\definecolor{ColorTitle}{cmyk}{0,.88,.77,.40}

\newcommand{\AMS}{\textsf{AMS-02}}

\newcommand{\eg}{\textit{e.g.}}
\newcommand{\ie}{\textit{i.e.}}

\newcommand{\p}{\textsf{p}}
\newcommand{\He}{\textsf{He}}

\newcommand{\Be}{\textsf{Be}}

\newcommand{\C}{\textsf{C}}

\newcommand{\BC}{\textsf{B}/\textsf{C}}

\newcommand{\BeBe}{\ensuremath{{}^{10}\textsf{Be}{/}{}^{9}\textsf{Be}}}

\usepackage{verbatim}
\usepackage{marginnote}
\usepackage{multirow}
\let\oldmarginpar\marginpar
\renewcommand\marginpar[1]{\-\oldmarginpar[\raggedleft\footnotesize #1]%
{\raggedright\footnotesize\color{red} #1}}
\begin{document}
\title{A Genetic Algorithm for Astroparticle Physics Studies}
\author{Xiao-Lin Luo$^1$, Jie Feng$^2$ and Hong-Hao Zhang}
\email{zhh98@mail.sysu.edu.cn}
\affiliation{School of Physics, Sun Yat-sen University, Guangzhou 510275, China}
\let\thefootnote\relax\footnotetext{X. L. Luo and J. Feng contributed equally to this work.\\
~$^1$ Also at School of Physics and Astronomy, Sun Yat-sen University,
Guangzhou 510275, China\\
~$^2$ Moved to Massachusetts Institute of Technology (MIT), Cambridge, Massachusetts 02139, USA}

\begin{abstract}
Precision measurements of charged cosmic rays have recently been carried out by space-born (e.g. AMS-02), or ground experiments (e.g. HESS). These measured data are important for the studies of astro-physical phenomena, including supernova remnants, cosmic ray propagation, solar physics and dark matter. Those scenarios usually contain a number of free parameters that need to be adjusted by observed data.
Some techniques, such as Markov Chain Monte Carlo and MultiNest, are developed in order to solve the above problem.
However, it is usually required a computing farm to apply those tools.
In this paper, a genetic algorithm for finding the optimum parameters for cosmic ray injection and propagation is presented. We find that this algorithm gives us the same best fit results as the Markov Chain Monte Carlo but consuming less computing power by nearly 2 orders of magnitudes.

\bf{{Program summary}}\\
\it{Operating system:} Linux\\
Programming Language: C\\
Software Package: ROOT\\
Libraries: cmath, cstdio, cstdlib, ctime\\
Optional Software Package: DRAGON
\end{abstract}
\pacs{}
\maketitle

\section{Introduction}

Energy spectra of cosmic rays (CRs) are essential for the investigation of astro-physical phenomena in the universe. Nowadays, more independent measurements of these spectra have been or are going to be published by new-generation experiments in space or on balloon, such as the Alpha Magnetic Spectrometer (AMS-02), the Advanced Thin Ionization Calorimeter (ATIC-2), the Calorimetric Electron Telescope (CALET), the Cosmic Ray Energetics and Mass (CREAM), the DArk Matter Particle Explorer (DAMPE), the Fermi Large Area Telescope (Fermi-LAT) and
the Payload for Antimatter Matter Exploration and Light-Nuclei Astrophysics (PAMELA), and by those on ground, such as the High Energy Stereoscopic System (HESS), the Cherenkov Telescope Array (CTA) and the Large High Altitude Air Shower Observatory (LHAASO).

The origin of the CRs and their experiences in the Galaxy together affect their spectra measured within our solar system. Thus, these spectra contain information about CR  acceleration processes and their transport effect. On top of that, signal from extra sources, including nearby sources and dark matter, can be extracted from spectral data of elementary particles.

The latest data of proton and nuclei spectra have challenged our conventional models based on linear diffusive-shock-acceleration (DSA) with homogeneous propagation in the interstellar medium (ISM) \cite{Strong:2007nh, Serpico:2015caa}.
These traditional models usually contain at least six free parameters,
\ie{} the spectral index of primary particle injection, the halo size, the convection wind velocity,  the Alfv$\acute{e}$n velocity,  the diffusion coefficient normalization as well as its spectral index.

For instance, PAMELA \cite{Adriani:2011cu} reported a spectral hardening in proton and helium spectra at around 200 GeV/nucleon and the different spectral indices for proton and helium,
which have been confirmed later by AMS-02
\cite{Aguilar:2015ooa, Aguilar:2015ctt}.
This anomaly has also been partially observed by ATIC-2 \cite{Panov:2009ak} and CREAM \cite{Yoon:2011aa, Yoon:2017qjx} at high energy range.
More recently, AMS-02 discovered the same spectral feature in the fluxes of other primary (\ie, accelerated in Galactic sources) nuclei (including carbon and oxygen) \cite{Aguilar:2017hno}. Moreover, the fluxes of secondary (\ie, spalled) nuclei (including lithium, beryllium, and boron) have a more significant hardening than those of primary ones \cite{Aguilar:2018njt}.
These unexpected features stimulate investigation
by the theoretical astrophysics community,
including new mechanisms in the CR injection \cite{Vladimirov:2011rn, Khiali:2016zzp} or propagation processes \cite{Erlykin:2012dp, Tomassetti:2015mha, Genolini:2017dfb}.
These implementations introduce at least three more free parameters in the model.

The CR positron flux measured by PAMELA \cite{Adriani:2008zr} and \AMS{} \cite{Accardo:2014lma, Aguilar:2014mma} shows a significant excess above $\sim$30 GeV and a cutoff at $\sim$200 GeV. Moreover, the spectra of CR leptons with a softening behavior at $\sim$0.9 TeV have been measured by DAMPE \cite{Ambrosi:2017wek} and HESS \cite{Aharonian:2008aa},
and confirmed by CALET \cite{Adriani:2018ktz}. These exciting features have triggered models involving: (i) nearby sources \cite{Mertsch:2014poa}; (ii) pulsars \cite{Serpico:2011wg, Yin:2008bs, Feng:2015uta, Grasso:2009ma, Hooper:2008kg, Mertsch:2010fn}; or (iii) dark matter \cite{Cholis:2008hb, Boudaud:2014dta,Yuan:2013eja,DiMauro:2015jxa,Feng:2017tnz}.
These models usually contain at least two free parameters.

Time variation of CR fluxes can also be a tool for studying solar physics.
Solar modulation of CRs can be described by the Parker's transport equation \cite{PARKER19659}.
Adopting Force-Field approximation \cite{Potgieter:2013pdj}, one can derive an analytical solution from that equation, which contains only one free parameter, \ie{} the solar modulation potential.
However, the accuracy of this approximation can not satisfy the precision of the recent CR data.
Particle drifts and adiabatic losses need to be taken into account in the models in order to explain the new phenomena,
\eg{} mass and charge-sign dependent modulation \cite{Adriani:2013as, Adriani:2016uhu}.
Some numerical solutions \cite{Boschini:2017fxq, Vittino:2017fuh,Kappl:2015hxv} have been proposed and usually involve more than ten free parameters.
The computations of those solar modulation models are very CPU/time consuming.
One can reconstruct local interstellar spectra of CRs with these numerical packages in order to study the CR diffusion in the Galaxy, especially in the energy range below 10 GeV.

To explain all the charged CR spectra simultaneously, one should scan the parameters in the multi-dimensional space to find the maximum likelihood.
In reality, it is not possible to derive an analytical solution for this maximum likelihood estimation, because the model contains too many parameters \cite{myung2003tutorial}
and the solutions in realistic model of CR propagation are numerical.
Fortunately, the problem can be solved numerically by optimization algorithms.
Recent work has proved that a bayesian analysis can be a powerful tool to achieve this goal.
In particular, Markov Chain Monte Carlo (MCMC) \cite{Putze:2010zn, Maurin:2011, Liu:2011re, Trotta:2011, Feng:2016loc} and Multi-Nest \cite{Johannesson:2016} are the most popular techniques in practical applications.
With the inputs of priors, the probability distributions of the parameters before data are taken into account,
and observations of the system, bayesian inference gives us the outputs of the posteriors, the conditional probability distributions of the parameters with the given data.
In reality, the bayesian analyses on the CRs require a lot of computing resources.
In this work, we propose an alternative technique, a genetic algorithm \cite{Goldberg1989Genetic, Haupt2004Practical, Kroese2011Handbook}, to obtain the optimal parameters.
Genetic algorithms give best fit results compatible with MCMC while consuming much less computing power.
It should be noted that the genetic algorithm does not provide the parameter uncertainties, and that it cannot fully replace but is complementary to MCMC.

This paper is organized as follows.
In Sect.~\ref{calcula} , we introduce the CR propagation model and show the details of our genetic algorithm.
In Sect.~\ref{result}, the best fit results about the propagation parameters are shown, and a comparison with the MCMC method is offered in terms of
the match with the experimental data and the computational efficiency.
We summarize our results in Sect.~\ref{conclu}.
\section{Calculations}\label{calcula}
\subsection{Cosmic Ray Propagation Model}\label{sub_propagation}
We consider that the propagation of all CR species in a two-dimensional model follows the transport equation with boundary conditions at $r=r_{max}$ and $z=\pm L$ as:
\begin{equation}
\frac{\partial\psi}{\partial t} = Q+\vec{\nabla}\cdot(D\vec{\nabla}\psi)-\psi\Lambda+\frac{\partial}{\partial E}(\dot{E}\psi),
\label{propagation_function}
\end{equation}
where $\psi=\psi (E,r,z)$ is the the particle number density as a function of energy and space coordinates.
The source term $Q$ contains a primary term and a secondary production term as $Q_{pri}$ and $Q_{sec}=\sum_j \Lambda_j^{sp}\psi_j$, the latter from spallation of heavier $j$-type nuclei with rate $\Lambda_j^{sp}$.
$\Lambda=\beta cn \sigma$ is the destruction rate for collisions of gas nuclei with density $n$
at velocity $\beta c$ and cross section $\sigma$;
$c$ is the speed of light and $\beta=v/c$ is the velocity of the particle $v$ divided by the speed of light.
The term $\dot{E}=-\frac{dE}{dt}$ describes ionization and Coulomb energy loss, as well as radiative cooling of CR leptons.
The spatial diffusion coefficient $D$ in the cylindrical coordinate system $(r,z)$, with radius $R_C$ of $\sim$ 20 kpc and half-height $L$, can be parameterized as
\begin{equation}
   D(\rho,r,z)=D_0(z)\beta^{\eta}\left(\frac{\rho}{\rho_0}\right)^{\delta(z)}\,,
\end{equation}
where $\rho\equiv pc/(Ze)$ is defined as the particle magnetic rigidity,
proportional to the particle momentum $p$ and the inverse of particle charge $Ze$.
$\delta(z)$ is the index of the power-law dependence of the diffusion coefficient on the rigidity.
$D_0(z)$ is the normalization of the diffusion coefficient at the reference rigidity $\rho=0.25~$GV.
In the spatial-dependent propagation model \cite{Tomassetti:2012ga, Tomassetti:2015mha, Feng:2016loc} adopted in this paper,
$\delta$(z) equals $\delta_0$ in the region of $|z|<\xi L$ (inner halo) and $\delta_0 + \triangle$ when $|z|\geq\xi L$ (outer halo).
Moreover, the normalization $D_0$ remains the same in the inner halo and becomes $\chi D_0$ in the outer halo.
There is a connecting function of the type $F(z) = (z/\xi L)^n$ for the smooth transition of the parameters $\chi$ and $\triangle$ across the two zones.
The exponent $\eta$ is set to be -4 in order to reproduce proton and nuclei spectra in the energy range below 20 GV.
The injection spectral indices of all the nuclei whose $z>1$ all equal to $\upsilon$ while that of proton is $\upsilon + \triangle\upsilon$.

In summary, the free parameters are
$D_0$, $\xi$, $L$, $\chi$, $\delta$, $\Delta$, $\nu$ and $\Delta\nu$ in this work.
They can be computed numerically with the propagation function (\ie{} Eq.\ref{propagation_function}) and the observed CR data (\ie{} cosmic ray proton (\p{}), helium (\He{}) and carbon (\C{}) fluxes, and Boron-to-Carbon (\BC{}) and Beryllium-10-to-Beryllium-9 ($^{10}\Be/^9\Be{}$) flux ratios).
Technically, the solution is obtained from the DRAGON code \cite{Gaggero:2013rya}, which is similar to GALPROP \cite{Boschini:2017fxq}. It takes around 150 seconds for a run in an 8-thread machine.
In this paper, the propagation parameters are computed with a Genetic Algorithm, which will be introduced in next section.

\subsection{Genetic Algorithm}\label{subsga}
\begin{figure}[!t]\centering
\includegraphics[width=7cm]{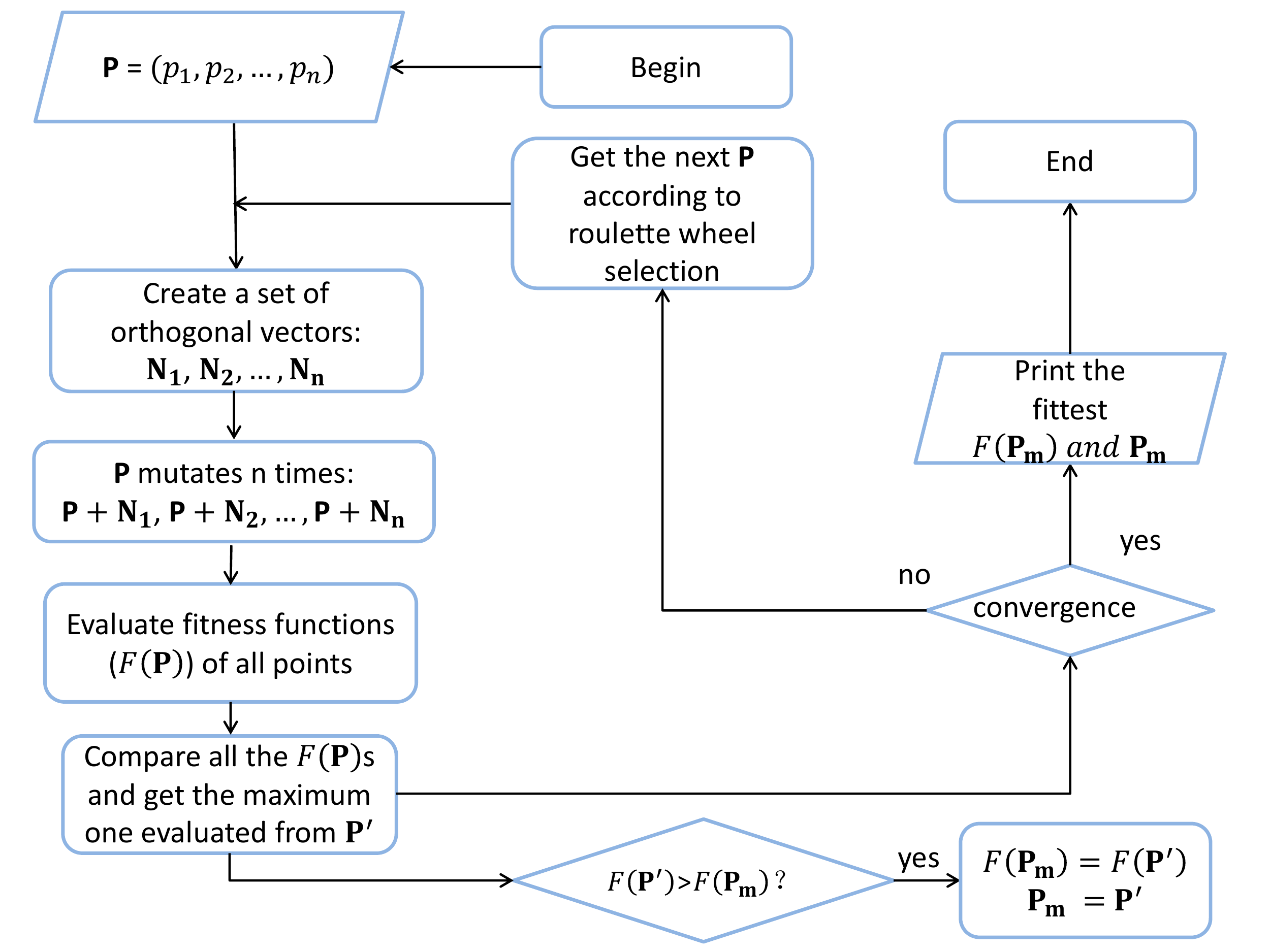}
\caption{The flowchart of the genetic algorithm.}\label{Fig:GA}
\end{figure}

The Genetic Algorithm is often used in optimization or search problems to generate high-quality solutions even in a complex parameter space.
This algorithm, as its name would suggest, is composed of processes of mutation and selection.
One generation involves one mutation process and one selection process.
It gives an optimized solution after some generations.
To simplify our problem, we would like to discuss the scenario with independent parameters first and leave the problem with correlated parameters at the end of this section. As is shown in Fig.\ref{Fig:GA}, the algorithm can be expressed in the following steps:
\begin{enumerate}[step 1]
\item The given parameter set $\bf{P}$ mutates into $n$ vectors as the first population.
\item The fitness functions $F(\bf{P})$s of all the vectors are evaluated.
\item $F(\bf{P})$s are compared with each other and with the maximum one, $F(\bf{P_m})$. The updated $F(\bf{P_m})$ is stored.
\item If a stopping criterion is met, the process is stopped; otherwise the process is repeated from step 1.
\end{enumerate}

The fitness function, equivalent to the likelihood function of the MCMC method, in Fig.\ref{Fig:GA} is defined as,
\begin{equation}\label{Eq::fitness}
F(\bf{P})=e^{-\frac{1}{2}\chi^2\bf(P)},
\end{equation}
where the $\chi^2\bf(P)$ describes the difference between experimental data and theoretical calculations from parameter set $\bf{P}$,
\begin{equation}\label{Eq::chisquare2}
\chi^2\left(\bf{P}\right)=
\sum_{k=1}^{N_{D}} \left( \frac{ y_{k}^{\rm exp} -
y_{k}^{\rm th}(\bf{P})}{\sigma_{k}} \right)^2.
\end{equation}
In Eq.\ref{Eq::chisquare2}, $k$ is the iterator for going through the $N_{D}$ data points,
while $y_{k}^{\rm exp}$, $y_{k}^{\rm th}(\bf{P})$ and $\sigma_{k}$ are the experimental value, theoretical prediction and uncertainty of the $k$th data point (with the consideration of solar modulation uncertainty
introduced by the Force-Field approximation \cite{Potgieter:2013pdj}) respectively.
\begin{figure}[!t]\centering
\includegraphics[width=7cm]{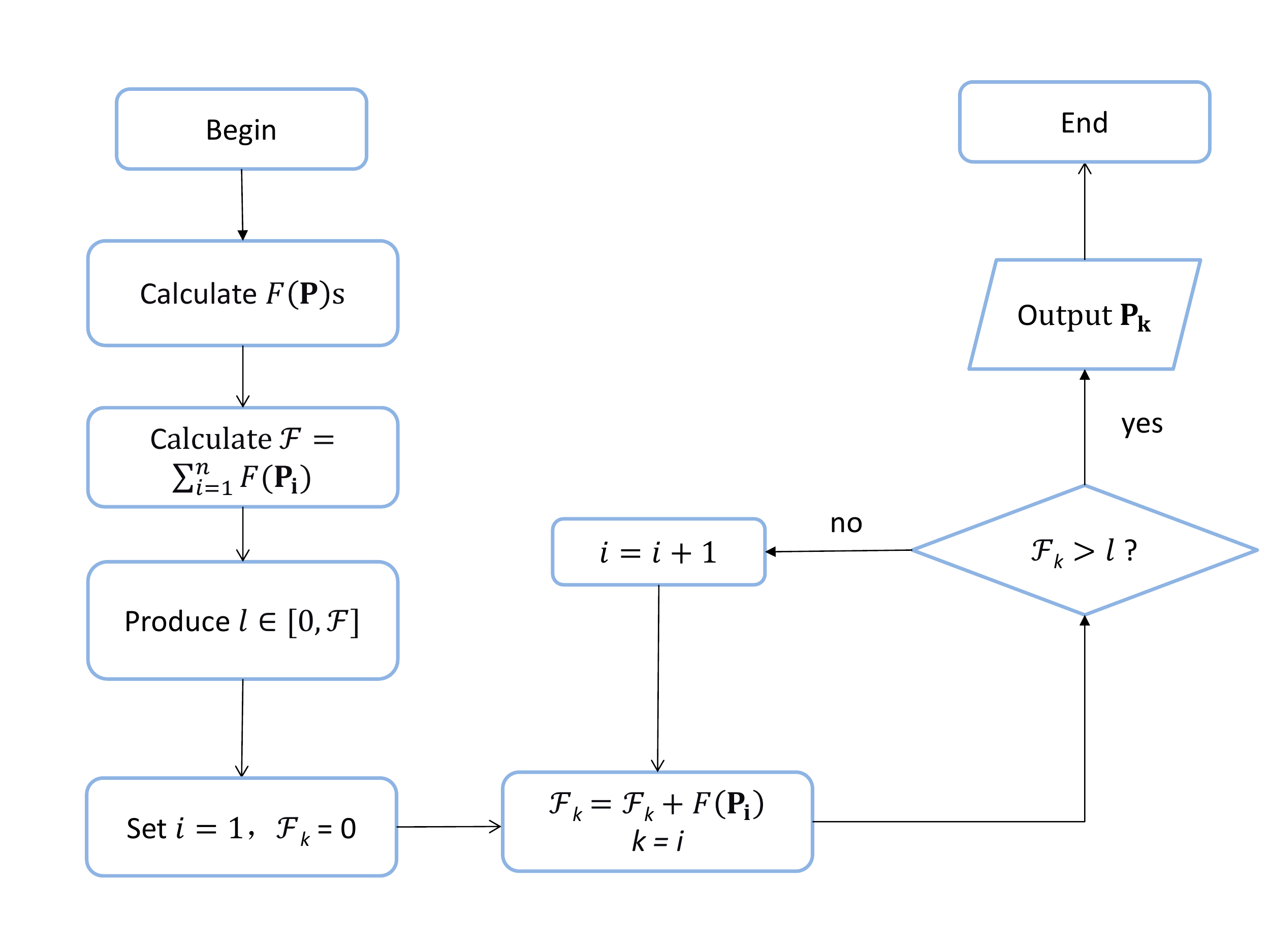}
\caption{The flowchart of the roulette wheel selection}\label{rw}
\end{figure}

The stopping criterion can be defined according to the practical problem.
In our case, it is sufficient to stop the process when it can not find a better solution after $8n$ (=64 in this work) generations.

\subsubsection{The mutation}\label{subsect_mutation}
As one of the important steps, the mutation process generates $n$ vectors based on the current $\bf{P}$ in the $n$-dimensional parameter space.
In the case with independent parameters, the mutations of $\bf{P}$, \ie{} $\bf{N}_{i} = \bf{P}_{i}-\bf{P}$, are orthogonal in order to obtain the highest efficiency of the scan.
Technically, the mutation process can be explained as the following steps:
\begin{enumerate}[step a]
\item A random number $|\bf{N}|$ is generated.
\item A set of $n$-dimensional unit orthogonal basis, $\bf{I_{i}}$ for the $i$th vector, is built by Gram-Schmidt process.
\item Let $\bf{N}_{i} = |\bf{N}|\cdot\bf{I_{i}}$.
\item  The $i$th vector is generated as $\bf{P}_{i} = \bf{P} + \bf{N}_{i}$.
\end{enumerate}

\subsubsection{The selection}
In case the process falls into a local maximum point, the roulette wheel selection is used to select $\bf{P}$ for the next generation. The probability to select one vector is proportional to its fitness function.
As is shown in Fig.\ref{rw}, the steps of roulette wheel selection can be listed as follows:
\begin{enumerate}[step i]
\item The fitness function $F(\bf{P})$s of all the points are calculated according to Eq.\ref{Eq::fitness}.
\item The sum of all the fitness functions is
\begin{equation}
\mathcal{F}=\sum_{i=1}^{n} F(\bf{P_i}),
\end{equation}

\item A uniform random number $l\in[0,\mathcal{F}]$ is generated.
\item The sums of $F(\bf{P_i})$ up to i=k, $\mathcal{F}_k=\sum_{i=1}^{k} F(\bf{P_i})$, are calculated, where $k$ is looped from 0. When $\mathcal{F}_k$ is greater than $l$, the process is stopped and the $\bf{P_k}$ is selected.
\end{enumerate}

\subsubsection{Treatment of correlated parameters}
In the scenario with correlated parameters,
which is the actually relevant scenario in CR astrophysics,
the mutation process discussed in Sect.~\ref{subsect_mutation} is not efficient.
This problem, however, can be easily transformed into a problem with independent parameters, which we have discussed in Sect.~\ref{subsect_mutation}.
To quantitatively describe the linear correlations between each two variables, we define the covariance matrix $\bf{M}$ as,
\begin{equation}\label{cov}
\bf{M_{ij}}\equiv \overline{\it{X_i X_j}},
\end{equation}
where $\bf{X}$ is a vector noted by\\
\begin{equation}
\langle X| = (p_1-\bar{p_1}, p_2-\bar{p_2}, \cdot\cdot\cdot, p_n-\bar{p_n} ),
\end{equation}
with an n-dimension random parameter $\langle p| = (p_1 \cdot\cdot\cdot p_n)$. \\
Thus, $X_i$ denotes $p_i-\bar{p_i}$ in Eq.~\ref{cov}.
If $\bf{M}$ is diagonal, the parameters are not linearly correlated.
Otherwise, we need to find the vector $\bf{X'}$ satisfying $\overline{X_i' X_j'}= a_i \delta_{ij}$, where $a_i=\bf{X'_{ii}}$ is the non-zero element of $\bf{X'}$, and $\delta_{ij}$ is the Kronecker delta.\\

We define a matrix $\bf{U}$ that satisfies
  \begin{equation}
  |X'\rangle =\bf{U} |\it{X}\rangle.
  \end{equation}
We can easily get
  \begin{equation}
  \overline{X_i' X_j'}=\overline{U_{il}X_l U_{jm}X_m}=U_{il}\overline{X_l X_m} U_{mj}^T.
  \end{equation}
If we define a diagonal matrix $\bf{D}$, we find
  \begin{equation}
  a_i \delta_{ij}=U_{il}\overline{X_l X_m} U_{mj}^T
  \end{equation}
  and
  \begin{equation}
  \bf{D}=\bf{UMU}^T.
  \end{equation}
We need to find the eigenvectors of $\bf{M}$ and combine them to get $\bf{U}$ and $\bf{X'}$ \cite{Liesen:2112873}.

In summary, we can make use of the prior to set $\langle\bar{p}| = (\bar{p_{1}} \cdot\cdot\cdot \bar{p_{n}})$ and some test runs to get the covariance matrix $\bf{M}$ as the first step.
We rotate the parameter space to diagonalize $\bf{M}$ as the second step. After that, we execute the parameter scan in the new space with the Genetic Algorithm.
If we do not have the covariance matrix in the first place to rotate the parameter space, the Genetic Algorithm works less efficiently.

\section{Results}\label{result}
\begin{table}[!t]
  \begin{tabular}{c | c | c | c c c}
    \hline\hline
     \multirow{2}*{\textbf{p}}& \multirow{2}*{\textbf{unit}}& $\>\>\>$\textbf{GA} & \multicolumn{3}{ c }{\textbf{MCMC}} \tabularnewline
    ~ & ~ & \textbf{best-fit} & $\>\>\>$\textbf{best-fit} & $\>\>\>$\textbf{$1 \sigma$-low} & $\>\>\>$\textbf{$1 \sigma$-up}  \tabularnewline
    \hline\hline
    $L$ & kpc  &  6.70 &  6.70 & \dots & \dots \tabularnewline
    \hline
    $D_{0}$ & 10$^{28}$\,cm$^{2}$\,s$^{-1}$  & 2.22 & 2.18 & 1.84 & 2.87 \tabularnewline
    \hline
    $\delta$ & \dots & 0.17 & 0.19&  0.12 & 0.23  \tabularnewline
    \hline
    $\Delta$ & \dots & 0.55 & 0.56 &  0.51 & 0.68 \tabularnewline
    \hline
    $\xi$ & \dots & 0.17 & 0.22 & 0.15 & 0.33  \tabularnewline
    \hline
    $\chi$ & \dots & 0.33 & 0.30 & 0.21 & 0.51 \tabularnewline
    \hline
    $\Delta\nu$ & \dots & 0.104 & 0.096& 0.090 & 0.119 \tabularnewline
      \hline
      $\nu$ & \dots & 2.27 & 2.29 & 2.20 & 2.40 \tabularnewline
      \hline\hline
  \end{tabular}
  \caption{\captionsize%
    Results of the Genetic Algorithm (GA) and MCMC scans for the transport and injection parameters (p) in terms of best-fit values for GA; and best-fit values and their bounds for $1-\sigma$ fiducial ranges (68\% C.L.) for MCMC.}
  \label{Tab::Posteriors}%
\end{table}

\begin{figure}[!t]
\includegraphics[width=0.5\textwidth]{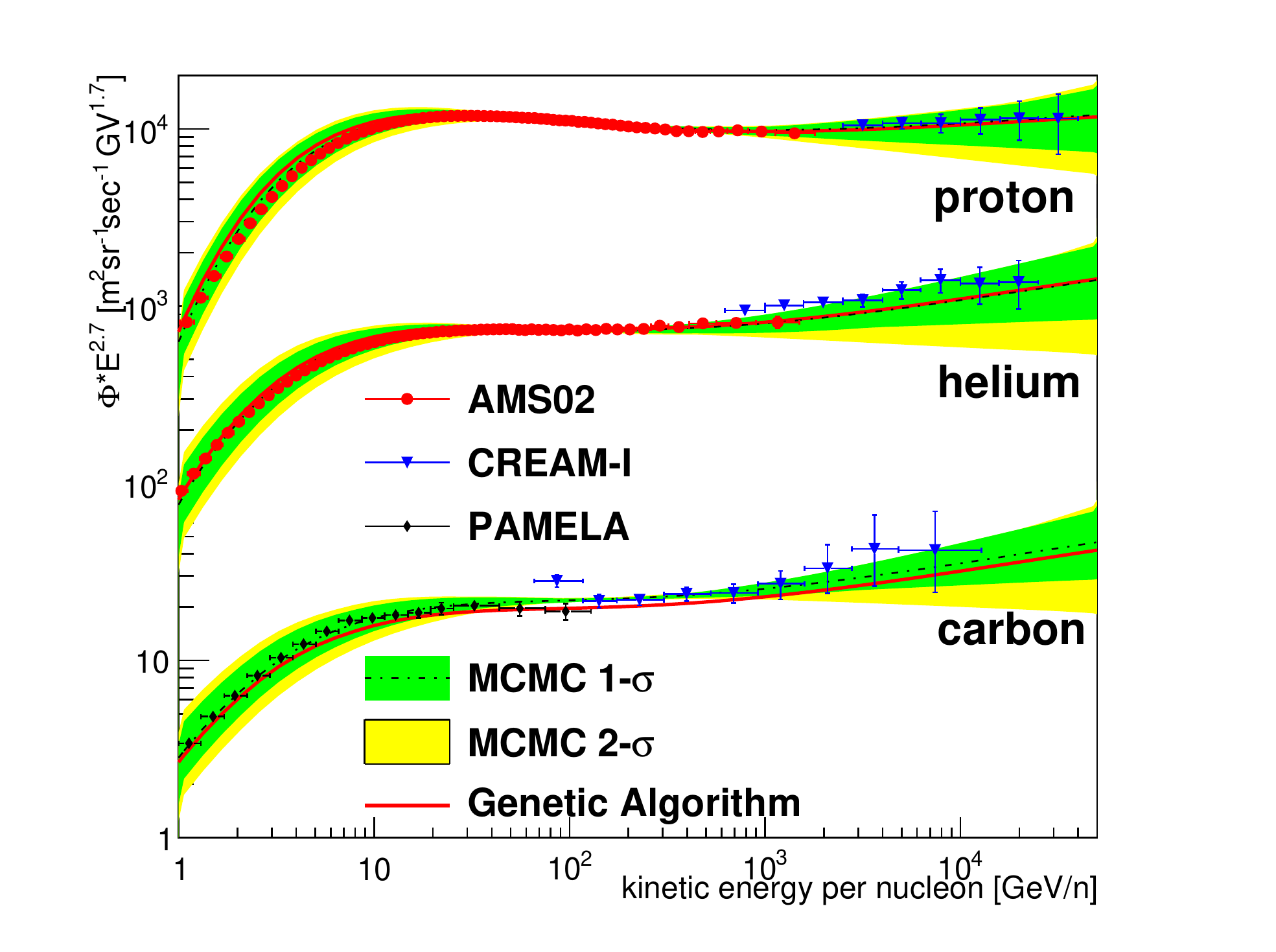} 
\caption{
Model prediction for proton, \He\ and \C\ fluxes compared with the experimental data
	  \cite{Aguilar:2015ooa, Yoon:2011aa, Aguilar:2015ctt, Adriani:2014xoa, Ahn:2008ApJ}. The black dash line is the best fit of MCMC, while the red solid line is that of Genetic Algorithm.
}
\label{Fig::ccProtonSpectrum}%
\end{figure}
\begin{figure}[!t]
\includegraphics[width=0.5\textwidth]{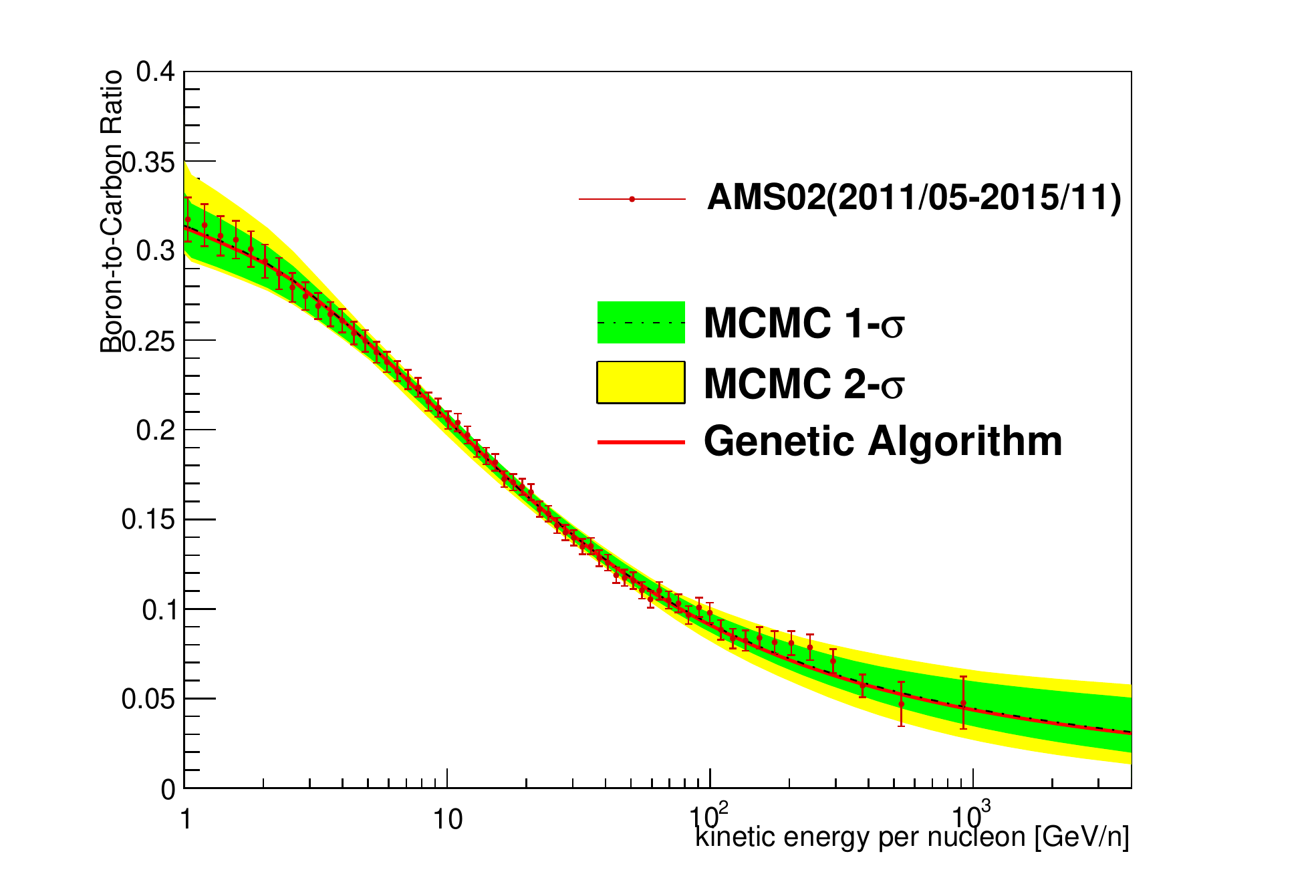}
\includegraphics[width=0.5\textwidth]{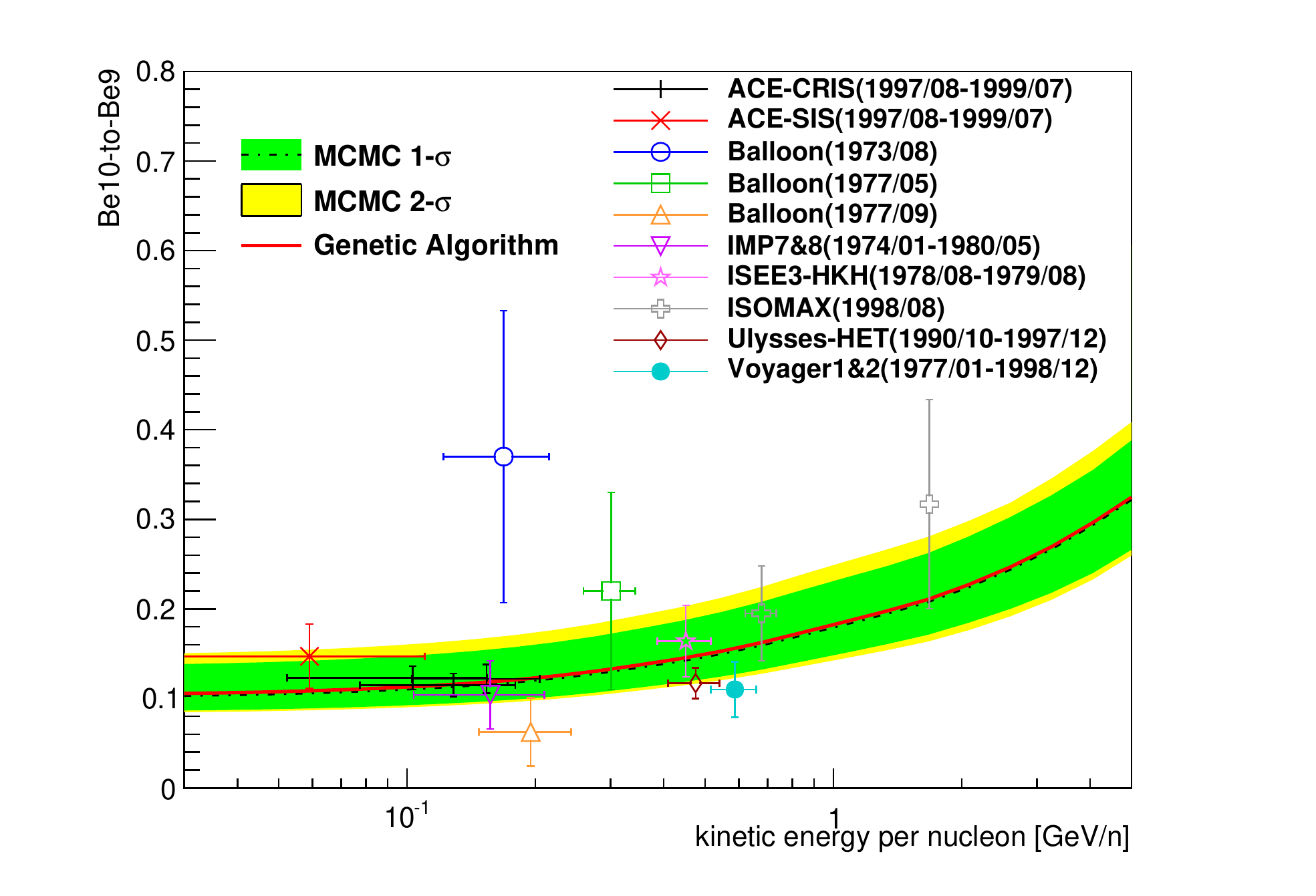}
\caption{
Top: Prediction of \BC\ compared with data \cite{Adriani:2014xoa, Panov:2007fe, Ahn:2008my, Obermeier:2012vg}.
Bottom: Prediction of \BeBe\ compared with data \cite{Hagen:1977ApJ,Buffington:1978ApJ,
Webber:1979ICRC, Garcia-Munoz:1977ApJ, Garcia-Munoz:1981ICRC, Wiedenbeck:1980ApJ, Connell:1998ApJ, Lukasiak:1994ApJ, Lukasiak:1997ICRC,
Lukasiak:1999ICRC}. The black dash line is the best fit of MCMC, while the red solid line is that of Genetic Algorithm.
}
\label{Fig::ccRatiosSpectrum}%
\end{figure}
\begin{figure}[!t]
\includegraphics[width=0.5\textwidth]{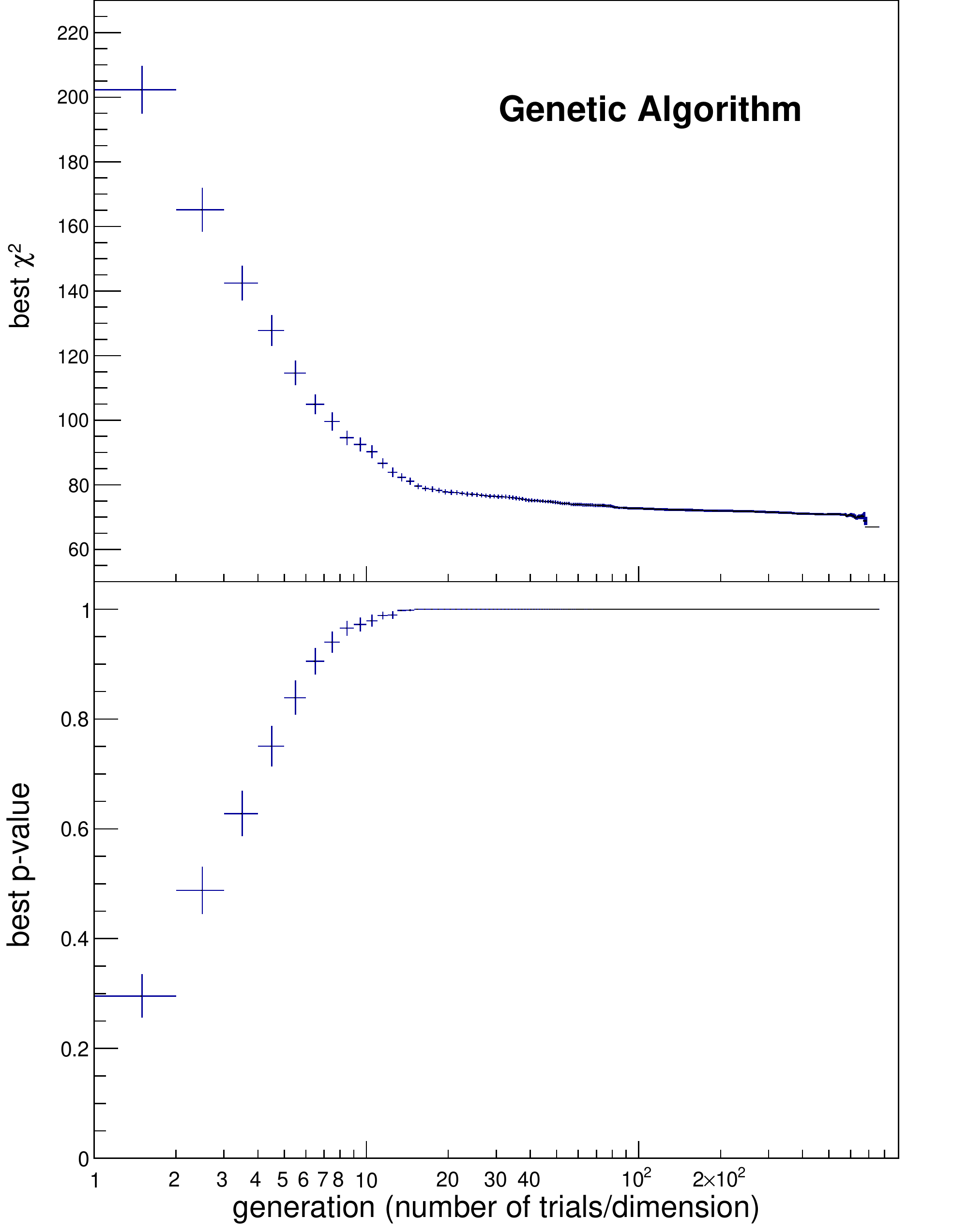}
\caption{
	Best $\chi^2$ (top) and p-value (bottom) change with number of generations in the Genetic Algorithm calculations. In this case, the number of dimension is 8.
}
\label{Fig::chi2_gen}
\end{figure}
\begin{figure}[!t]
\includegraphics[width=0.5\textwidth]{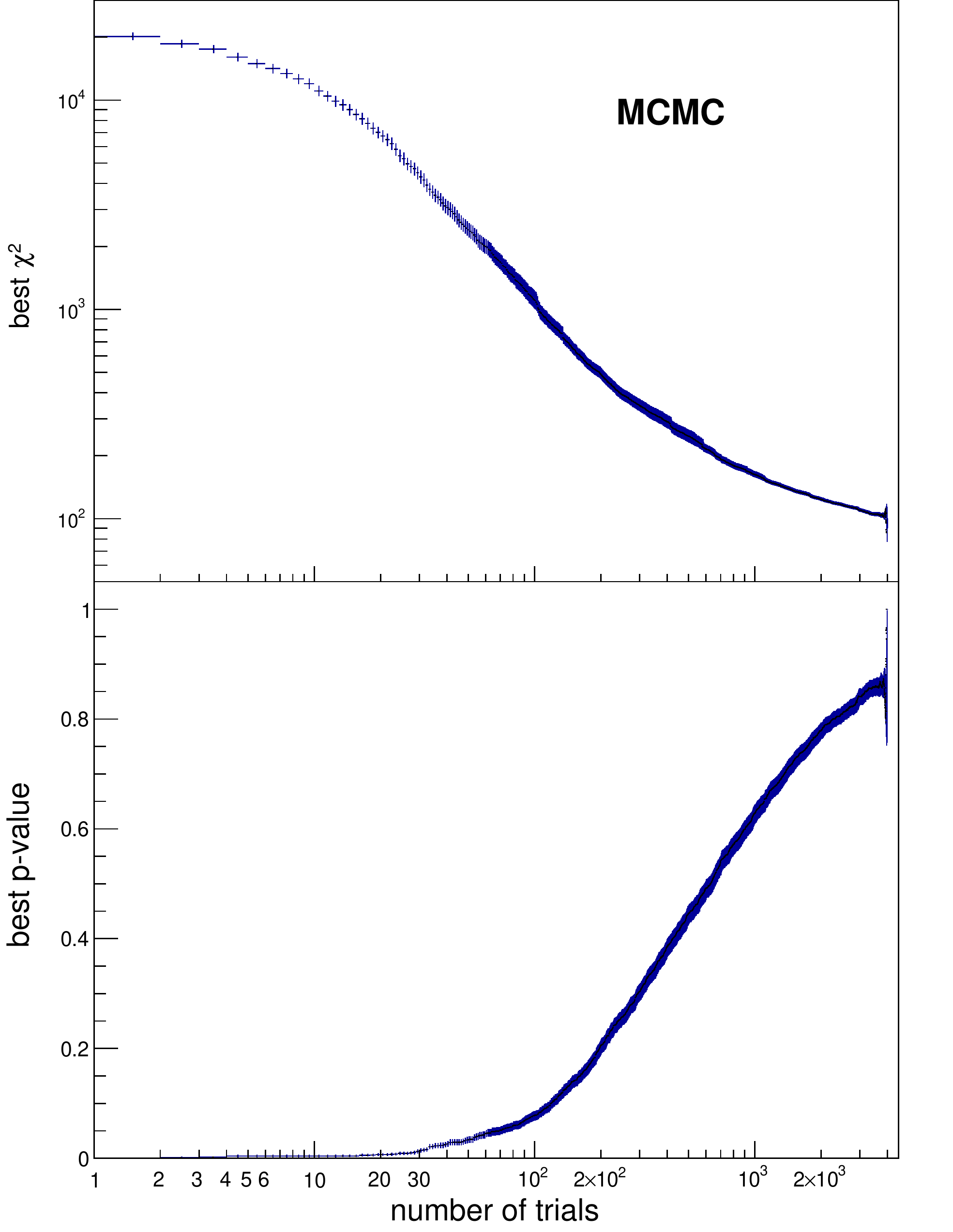}
\caption{
	Best $\chi^2$ (top) and p-value (bottom) change with number of trials in the Markov Chain Monte Carlo calculations.
}
\label{Fig::chi2_step_mcmc}
\end{figure}
In this section, we compare the Genetic Algorithm with the MCMC method.
As mentioned in Sect.~\ref{sub_propagation}, the fit is operated in a 8-dimension space with the  degree of freedom of 155. We take the best fit parameters in Ref.~\cite{Feng:2016loc}, where AMS-02 \BC{} data were not included, as the priors.
What we are interested in are the bias of those results as well as the corresponding efficiencies.
We repeated the Genetic Algorithm calculation 100 times and the calculation by MCMC 700 times, in order to estimate their average performance.

As expected, Fig.~\ref{Fig::ccProtonSpectrum}, Fig.~\ref{Fig::ccRatiosSpectrum} and Tab.~\ref{Tab::Posteriors} show that the best fit result obtained by the Genetic Algorithm is consistent with those obtained by the MCMC method.
The best fit results are the maximum likelihood results found by these two methods.
These plots show that there is no bias in the Genetic Algorithm.

In order to quantify the computing speed, one can look into the goodness of the model as a function of generations in Fig.~\ref{Fig::chi2_gen}. In our case, 8 trials are performed by one generation. The errors show the standard errors on the mean values.
The best $\chi^2$ drops and its corresponding p-value, defined as the cumulative distribution function value for the chi-square distributions for the given number of degree of freedom (=155), rises as generation increases. To make a fair comparison, we plot the same values as a function of number of trials in the MCMC process in Fig.~\ref{Fig::chi2_step_mcmc}.
On average, the Genetic Algorithm finds out the best p-value close to 1 after 15 generations, which is equivalent to $15\times8 = 120$ trials, as is shown in Fig.~\ref{Fig::chi2_gen}. On the other hand, Fig.~\ref{Fig::chi2_step_mcmc} tells us that it takes ~3000 trials for the MCMC process to find the result with a p-value greater than 0.9 on average.
The Genetic Algorithm is faster than the MCMC process by a factor of 70 to achieve p-value greater than 0.9.
This computing speed estimation shows that the Genetic Algorithm is in advance for the optimization problem compared with the MCMC process.

However, the Genetic Algorithm cannot provide the uncertainty estimation of parameters as MCMC does.
The uncertainties are useful to describe the significance of the model.
To overcome this limitation, one may get the best fit parameters with the Genetic Algorithm first, and then set them as the prior to MCMC to estimate the uncertainties of the parameters.

\section{Conclusion}\label{conclu}
This work is aimed at solving the best-parameter-finding problem using less computing power.
A  Genetic Algorithm has been developed to achieve this goal.
In particular, we show that the Genetic Algorithm works quite well for CR propagation studies.
Compared with the existing popular tools, MCMC for instance, the Genetic Algorithm gives fairly good results but 70 times faster for the same initial condition.

The disadvantage of the Genetic Algorithm is that the uncertainties of the parameters can not be retrieved.
If the uncertainties are of interest, one can estimate them with a MCMC process with priors given by the best fit of the Genetic Algorithm.
This combination of the two tools will give the same performance while requiring less computing power compared to MCMC alone.

The code is currently written in C language and will be released in public soon. It will be useful for astro-particle physics studies with multiple parameters.

\section*{Acknowledgments}
This work is
supported by the National Natural Science Foundation of China (NSFC) under Grant No.
11875327, the Fundamental Research Funds for the Central
Universities, the Natural Science Foundation of Guangdong Province under Grant No.
2016A030313313, and the Sun Yat-Sen University Science Foundation.
\bibliography{ga_script}
\appendix

\section{Description and Manual of the Code}
This document is about how to run the Genetic AlgoriThm for cOsmic Ray studies (GATOR). 
The source code of GATOR is in C. 
This example requires ROOT package to be installed (installation see https://root.cern.ch/downloading-root) in Linux operating systems.
ROOT is used to generate random variables and is not mandatory. \\
\\

\subsection{Prerequisite for the Code}
The program contains three files: the main program ``GA0.C'', the chi-square calculation function ``test\_getchi2.h'', a main program script ``main.C'' and a ``Makefile''. 
\subsection{Description of the chi-square calculation file}
The file ``test\_getchi2.h'' contains a function, which returns the chi-square value for a given set of input parameters. 
For cosmic ray studies, the chi-square is a value to describe how the model matches the observed data. 
In this document, an example containing 8 parameters is defined as :
\begin{equation} \label{Eq::chisquare2}
\chi^2\left(\bf{P}\right)=
\sum_{k=1}^{8} \left( \frac{ p_{k} -
p_{k}^{BestFit}}{\sigma_{k}} \right)^2,
\end{equation}
where the best fit values $\bf{p^{BestFit}}$ and sigmas $\bf{\sigma}$ are:\\
\\
$\bf{p^{BestFit}}$ = (6.69891, 2.17953, 0.189886, 0.559855, 0.216936, 0.303952, 0.0959849, 2.29343),\\
$\bf{\sigma}$ = (2, 1.2, 0.15, 0.15, 0.15, 0.15, 0.04, 0.15).\\

It will be good to print out the results at the end of this function. In this example, it is written that:
\begin{verbatim}

cout << "Chisqure: " << chi2 << " 155 " ;
for (int i = 0;i<np;i++) cout << p[i] << " " ;
cout << endl;

\end{verbatim}

\subsubsection{Particular Example for DRAGON}
In this paper, the cosmic ray transport code DRAGON (https://github.com/cosmicrays) is used compute the model predictions. 
GALPROP (https://galprop.stanford.edu/ )  is an alternative code to obtain the model predictions. 
The corresponding chi-square calculation function is in the file ``getchi2.h''. 
If one wants to adopt this function, he should replace:
\begin{verbatim}
#include "test_getchi2.h"
\end{verbatim}
 in the file ``GA0.C'' by 
 \begin{verbatim}
#include "getchi2.h"  .
\end{verbatim}

In the file ``getchi2.h'', there are basically two steps: execution of the model computation with DRAGON and calculation of the chi-squares. 
In C, the function "std::system" allows us to execute external program via a bash script ``bc\_par\_Be.sh'' with the input parameters. 
This bash script ``bc\_par\_Be.sh'' creates an input file according to the parameters and then execute DRAGON. 
The definitions of the chi-squares can be found in Sect.(Add reference).

\subsection{Execution}
One needs to give the initial values, the expected widths, the lower limits and the upper limits to GATOR, for instance:\\
\begin{verbatim}

double p[8]={6.71, 1.83, 0.18, 0.58, 
             0.19, 0.42, 0.096, 2.30};
double sigma_p[8]={2, 1.2, 0.15, 0.15, 
                 0.15, 0.15, 0.04, 0.15};
double p_limitleft[8]={2.5, 0.5, 0., 0.2, 
                     0.08, 0.2, 0.03, 2.0}; 
double p_limitright[8]={9.5, 5.0, 0.6, 1.2, 
                    0.6, 1.2, 0.15, 2.6};

\end{verbatim}

The script ``main.C'' serves as an interface inputing those values to GATOR. It integrates the ROOT commands:
\\
\begin{verbatim}

GA0(p,sigma_p,8,p_limitleft,p_limitright);

\end{verbatim}

To compile this program, one should use the command:
\begin{verbatim}

make

\end{verbatim}

It will generate an executable file. To execute this program, one should use the command:
\begin{verbatim}

./GA.exe > data txt

\end{verbatim}

\subsection{Extraction of the Results}
The results are stored in the file ``data.txt''. For instance, to extract the results, one could use the command:
\begin{verbatim}

cat data.txt | grep Chisqure

\end{verbatim}

After you get the results of all the trials, you can find the least chi-square one as well as the corresponding variables at the end of the output file.

\end{document}